\documentclass[preprint,showpacs,preprintnumbers,amsmath,amssymb]{revtex4}
\usepackage{graphicx}
\usepackage{graphicx,amssymb,amsmath}
\usepackage{graphics}
\usepackage{multirow}
\usepackage{caption}
\usepackage{subcaption}

\begin{document}

\title{Competing addition processes give distinct growth regimes in the assembly of 1D filaments}
\author{Sk Ashif Akram$^1$}
\author{Tyler Brown$^1$}
\author{Stephen Whitelam$^2$}
\author{Georg Meisl$^3$}
\author{Tuomas P.J. Knowles$^4$}
\author{Jeremy D. Schmit$^1$}
\email{schmit@phys.ksu.edu}
\affiliation{$^1$Department of Physics, Kansas State University, Manhattan, KS 66506, USA }
\affiliation{$^2$Molecular Foundry, Lawrence Berkeley National Laboratory, 1 Cyclotron Road, Berkeley, CA 94720, USA}
\affiliation{$^3$Yusuf Hamied Department of Chemistry, Centre for Misfolding Diseases, University of Cambridge, Cambridge, United Kingdom}
\affiliation{$^4$Cavendish Laboratory, Department of Physics, University of Cambridge, Cambridge, United Kingdom}

\begin{abstract}
We present a model to describe the concentration-dependent growth of protein filaments. Our model contains two states, a low entropy/high affinity ordered state and a high entropy/low affinity disordered state. Consistent with experiments, our model shows a diffusion-limited linear growth regime at low concentration, followed by a concentration independent plateau at intermediate concentrations, and rapid disordered precipitation at the highest concentrations. We show that growth in the linear and plateau regions is the result of two processes that compete amid the rapid binding and unbinding of non-specific states. The first process is the addition of ordered molecules during the periods where the end of the filament is free of incorrectly bound molecules. The second process is the capture of defects, which occurs when consecutive ordered additions occur on top of incorrectly bound molecules. We show that a key molecular property is the probability that a diffusive collision results in a correctly bound state. Small values of this probability suppress the defect capture growth mode, resulting in a plateau in the growth rate when incorrectly bound molecules become common enough to poison ordered growth. We show that conditions that non-specifically suppress or enhance intermolecular interactions, such as the addition of depletants or osmolytes, have opposite effects on the growth rate in the linear and plateau regimes. In the linear regime stronger interactions promote growth by reducing dissolution events, but in the plateau regime stronger interactions inhibit growth by stabilizing incorrectly bound molecules.
\end{abstract}

\maketitle

\section{Introduction}

Numerous biological and bio-mimetic processes depend on the formation of highly ordered 1D assemblies. Examples include essential cytoskeleton filaments as well as pathological assemblies such as amyloid fibrils and aggregates of sickle cell hemoglobin. In many of these cases the thermodynamic basis of interactions stabilizing the final state are known, but the kinetic factors that determine the assembly mechanism are less well understood. In particular, biomolecules have low symmetry, which necessitates a large search over rotational, conformational, and translational states to find the lowest free energy state \cite{Hall2005,Schmit2012,Schmit2013}. Furthermore, assembly processes occurring in vivo will be exposed to a heterogeneous mixture of molecules, each of which is capable of impeding assembly with non-specific interactions. Successful assembly requires the system to sample these non-specific states, which greatly outnumber the specific state, much faster than the observation timescale. This sampling process fails when the non-specific affinity increases the binding lifetime of incorrect states. This leads to a slowing of the assembly rate, known as self-poisoning in single component systems \cite{Ungar2005,deyoreo2003,Whitelam2015b}, and can even arrest the system in long lifetime traps.

In this paper we present a model for the growth of ordered 1D filaments to understand the self-assembly behavior of amyloid fibrils. Amyloids are 1D protein aggregates associated with numerous neurological disorders \cite{Hardy2002}. Bulk aggregation experiment reveal a complex interplay of multiple processes, including primary nucleation, fibril elongation, fragmentation, and secondary nucleation \cite{Michaels2023}. Here we focus exclusively on the elongation of existing fibrils. The elongation reaction shows two distinctive behaviors as a function of concentration. At low concentration, the growth rate increases linearly with concentration, indicating that diffusion is limiting. At higher concentration the growth rate plateaus and becomes nearly independent of concentration, which indicates that the sampling process is limiting \cite{Schmit2013,Buell2010}. While most aggregation experiments are conducted in the linear regime, the plateau regime is likely to be more informative about the mechanism of molecular reconfiguration. However, it is not known how to extract this information from growth curves.

Our model contains three parameters, which describe the affinity of both the specific and non-specific states, as well as the multiplicity of the non-specific ensemble. We show that fibril growth can be divided into two distinct processes that we call ``ordered addition'' and ``defect capture''. Ordered addition dominates at low concentration when the binding lifetime of non-specific states is short compared to the diffusion arrival time. At higher concentration ordered growth is progressively poisoned by non-specifically bound molecules. At the same time, it is increasingly likely that incorrectly bound molecules become captured in the fibril as defects. However, defect incorporation requires rare events where consecutive ordered addition events capture incorrectly bound molecules. These defect capture events are strongly suppressed when the probability of specific binding is low, which results in the appearance of a plateau, or even decline, in the growth rate. Thus, it is the competition between the two growth processes that determines the shape of the growth curve.

\section{Model}
\begin{figure}
\includegraphics[width=0.7\textwidth]{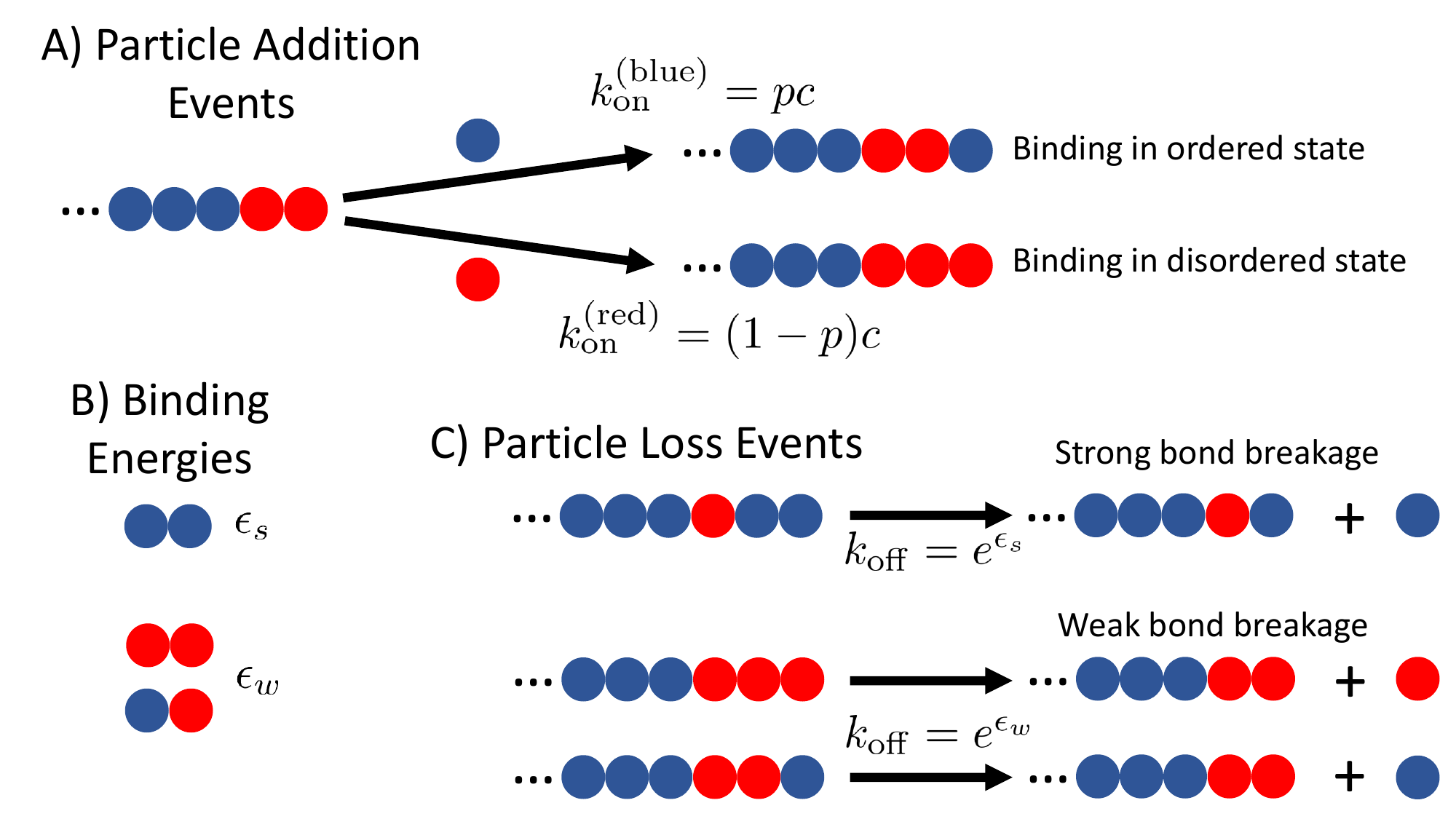}
\caption{The model is defined by three parameters: $p$, $\epsilon_s$, and $\epsilon_w$. A) Particles add to the end of the filament with a total rate proportional to the concentration $c$. Addition events can be in either the ordered state (blue), which occurs with probability $p$, or the disordered state (red), which occurs with probability $(1-p)$. B) Binding energies depend on the state of both neighboring particles. If both particles are blue the interaction is ``strong'' and takes a value $\epsilon_s$. If either of the particles are red the interaction is ``weak'' and has a value $\epsilon_w$. C) The last particle on the filament can detach with a rate that depends on its binding energy. }\label{FigModel}
\end{figure}

We consider a solution of particles that self-assemble to form one-dimensional filaments. In our model particles can bind to the end of the filament in one of two states  that we label blue and red \cite{Whitelam2012,Whitelam2014,Whitelam2016a}. The blue state is energetically favored and represents ordered molecules while the entropically favored red state represents the ensemble of disordered states. The model can describe either the self-assembly of particles in a heterogenous mixture, in which case the red particles represent bystander molecules, or the self-assembly of anisotropic molecules, in which case red particles represent molecules in incorrect orientations or alignments. Without loss of generality, we focus our discussion on the latter situation. Particles bind to the end of the filament at a total rate proportional to the concentration, $k_\mathrm{on}=k_\mathrm{diff}c$. Hereafter we set the diffusion collision rate constant to unity, $k_\mathrm{diff}=1$. The particle binding events can be divided into events leading to the blue state, which occurs with probability $p$, and red binding event occurring with probability $(1-p)$. Thus, the attachment rates are $k_\mathrm{on}^\mathrm{(blue)}=p c$ and $k_\mathrm{on}^\mathrm{(red)}=(1-p) c$ (Fig. \ref{FigModel}A). We do not consider the possibility of molecules converting between red and blue states while bound to the filament, which is an approximation that has empirical support in the modeling of protein assemblies \cite{Schmit2012,Schmit2013,Jia2017,Jia2020}.

While we do not consider fragmentation in the middle of a filament, the end particle of the filament can unbind and return to the solution state. The rate of unbinding depends on the binding energy $\epsilon_i$ between the terminal particle and the penultimate particle
\begin{equation}\label{eq:koff}
  k_\mathrm{off}=e^{\epsilon_i}
\end{equation}
The binding energies, which are given in units of $k_BT$, can take one of two values; if the last two particles are both blue the binding energy is ``strong'' $\epsilon_i=\epsilon_s$, whereas if either or both particles are red the energy takes a ``weak'' value $\epsilon_i=\epsilon_w$ (Fig. \ref{FigModel}B,C). Thus, the growth rate at a particular concentration is defined by the parameters $p$, $\epsilon_s$, and $\epsilon_w$.

\section{Results}

\begin{figure}
\includegraphics[width=0.5\textwidth]{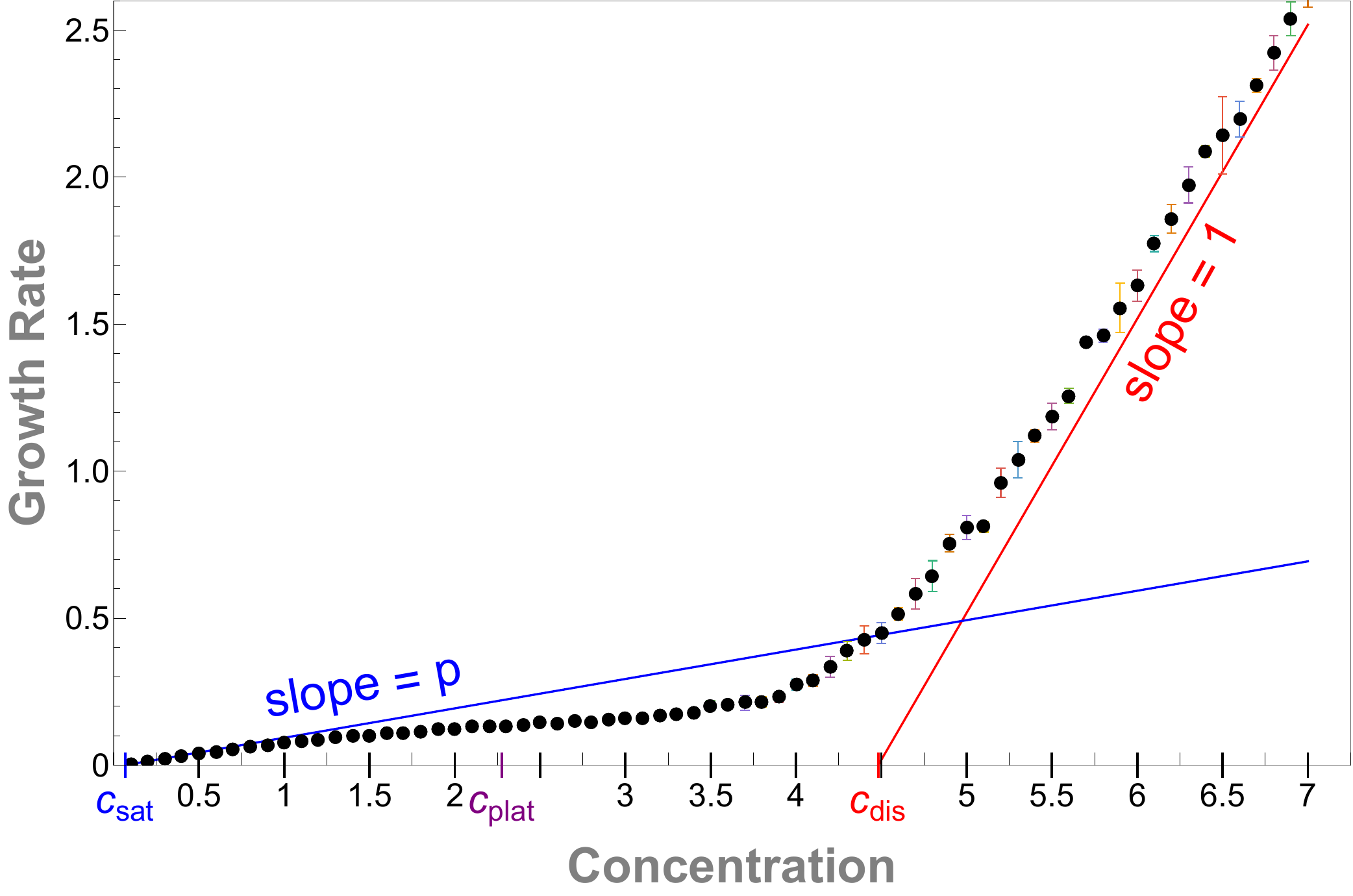}
\caption{Filament growth curves show two linear regimes separated by a plateau. The low concentration linear regime begins at the saturation concentration $c_\mathrm{sat}$ and the high concentration linear regime begins at $c_\mathrm{dis}$, which is the concentration where disordered growth becomes stable. Between these two regimes is the plateau, which is centered around a concentration $c_\mathrm{plat}$. $\epsilon_s= -5$, $\epsilon_w= 1.5$, $p=0.11$.}\label{FigRegimes}
\end{figure}

{\bf Ordered and disordered linear growth regimes occur at low and high concentrations, respectively.} Kinetic Monte Carlo simulations of the model described above reveal linear growth regimes at low and high concentrations separated by a ``self-poisoning'' plateau at intermediate concentrations (Fig. \ref{FigRegimes}). In the low concentration regime filament growth is dominated by the ordered state so that the net growth rate is $R\simeq pc - e^{\epsilon_s}$, where the two terms describe the attachment and detachment of blue particles. The solubility of the ordered state is determined from the condition that the growth rate is zero, which gives
\begin{equation}\label{eq:Csat}
  c_\mathrm{sat}=e^{\epsilon_s}/p
\end{equation}
At concentrations near this threshold, the grown fibrils will contain defects at a frequency that can be estimated from the two-state partition function $F_\mathrm{red}^\mathrm{(eq)}=\omega_r e^{-\epsilon_w}/(e^{-\epsilon_s}+\omega_r e^{-\epsilon_w})$, where $\omega_r=p^{-1}-1$ is the multiplicity of the red state.

In the high concentration limit particles deposit much faster than the disordered detachment rate, $e^{\epsilon_w}$, so the filament grows primarily by the addition of red particles. In this regime, the net growth rate is $R_\mathrm{dis}=c-e^{\epsilon_w}$. The condition $R_\mathrm{dis}=0$ provides a criteria for the onset of disordered aggregation
\begin{equation}\label{eq:Cdisorder}
  c_\mathrm{dis}=e^{\epsilon_w}
\end{equation}
Our model will break down for concentrations greater than $c_\mathrm{dis}$ because disordered aggregation will violate our assumption of one-dimensional assembly. This is because incorrectly bound molecules lack the orientational specificity to enforce unidirectional growth. In protein systems this will often result in $\beta$-rich, but non-fibrillar, amorphous aggregates.

\begin{figure}
\includegraphics[width=0.7\textwidth]{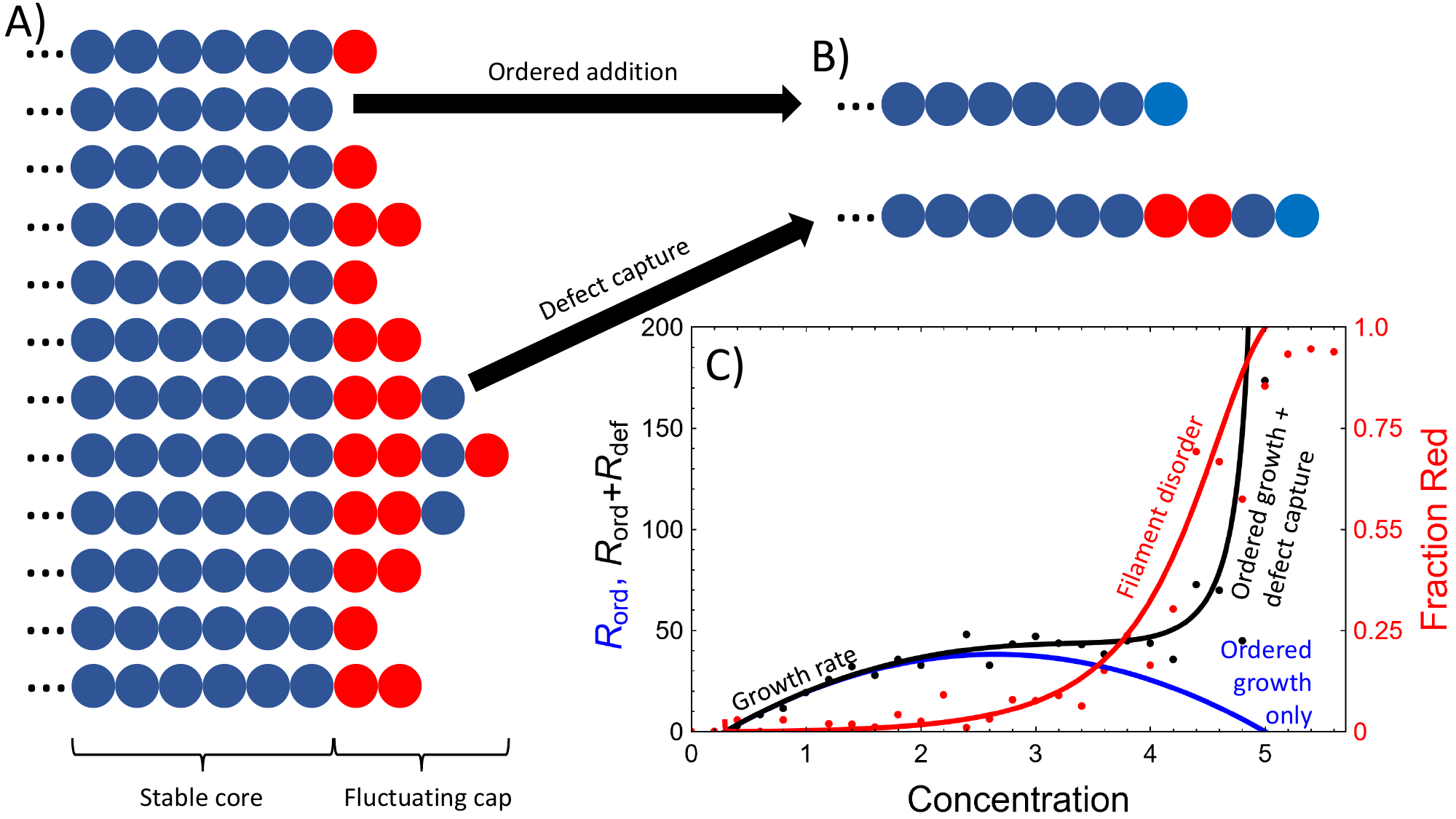}
\caption{ A) The filament can be modeled as a stable core, consisting mostly of blue particles, and a rapidly fluctuating disordered cap. The cap may contain blue particles provided they only have red neighbors such that all the interactions energies in the cap are $\epsilon_w$.  B) In the concentration regime below the point where disordered aggregation is stable ($c<c_\mathrm{dis}$), the filament growth occurs by two distinct processes. Ordered addition occurs when the disordered cap momentarily disappears, which allows a blue particle to bind directly to the ordered core. Defect capture occurs when two consecutive blue particles bind to the end of the disordered cap. This forms a strong bond that captures all previous red particles. C) In the low concentration regime filament growth is dominated by ordered addition (Eq. \ref{eq:rateordered}, blue line). In the plateau the contribution from ordered addition diminishes and defect capture takes over. The sum of ordered growth and defect capture (Eqs. \ref{eq:rateordered} and \ref{eq:ratedefect}, black line) is in good agreement with kinetic Monte Carlo simulations (black points). A consequence of the greater contribution of defect capture is that an increasing fraction of the filament is red (Eq. \ref{eq:fractionred}, red line/points).}\label{FigProcesses}
\end{figure}

{\bf Growth stagnates when disordered particles block ordered addition.} The most interesting behavior occurs in the plateau separating the ordered and disordered regimes. In this regime, the red state is not thermodynamically stable, in the sense that the net addition rate is negative. However, red addition events occur frequently enough to poison blue additions. The growth rate below the disorder transition $c<c_\mathrm{dis}$ can be understood as the sum of two processes that we refer to as ``ordered addition'' and ``defect capture'' \cite{Schmit2013}. These two processes are illustrated in Fig. \ref{FigProcesses}. Ordered addition can only occur at the times when there are no red particles bound at the end of the filament. Defect capture occurs when two consecutive blue particles add atop one or more red particles. The second blue particle results in a strong binding energy that effectively traps the red particles within the filament.

To estimate the contributions of these processes we compute the statistics of the disordered particles that form a cap at the end of the ordered filament. If $p\ll 1$ and the system is not too close to the disordered regime $c<c_\mathrm{dis}$, then many red attachment and detachment events occur between each blue addition. Under these conditions the disordered cap will nearly equilibrate in the time between the creation of blue-blue bonds. The grand partition function of the disordered cap is approximately given by
\begin{equation}\label{eq:capPartition}
  Q_\mathrm{cap}=\sum_{n=0}^{\infty}\left(c e^{-\epsilon_w} \right)^n=\frac{1}{1-c e^{-\epsilon_w}}
\end{equation}
where $c=e^\mu$ serves as the fugacity. This expression does not contain $p$ because both red and blue particles are allowed within the disordered cap provided that there are not two adjacent blue particles. Eq. \ref{eq:capPartition} is obtained by neglecting the possibility of consecutive blues, introduces an error of order $p$ (See Appendix for a more exact calculation). Defining $P_n$ as the probability the disordered cap contains $n$ particles, the probability the cap is absent is $P_0=1/Q_\mathrm{cap}$. Thus, the growth rate due to the ordered addition mechanism is given by
\begin{equation}\label{eq:rateordered}
  R_\mathrm{ord}=P_0(pc - e^{\epsilon_s})=(1-c e^{-\epsilon_w})(pc - e^{\epsilon_s})
\end{equation}
Note that $P_0$ multiplies both the addition and dissolution terms because the presence of a disordered cap inhibits both processes. Eq. \ref{eq:rateordered} is plotted as a blue line in Fig. \ref{FigProcesses}B. At low concentration it rises linearly with the rate $R=pc$, consistent with Fig. \ref{FigRegimes}, reflecting the fact that increasing concentration promotes more rapid particle additions. However, at higher concentration the term proportional to $c^2$ becomes dominant and $R_\mathrm{ord}$ decreases with concentration. This behavior is explained by the fact that at these concentrations it becomes increasingly unlikely for the disordered cap to disappear entirely. The non-monotonic growth rate predicted by Eq. \ref{eq:rateordered} is often hidden because growth from defect incorporation compensates for the decline in ordered additions at high concentration.

The maximum value of Eq. \ref{eq:rateordered} provides a useful estimate for the plateau height and the concentration at which the plateau begins. We define the plateau concentration by $d R_\mathrm{ord}/dc=0$, which gives
\begin{equation}\label{eq:plateauconcentration}
  c_\mathrm{plat}=\frac{e^{\epsilon_w}+e^{\epsilon_s}/p}{2}
\end{equation}
which is exactly halfway between the solubility thresholds for the ordered and disordered phases $c_\mathrm{ord}$ and $c_\mathrm{dis}$. The plateau height is obtained by inserting Eq. \ref{eq:plateauconcentration} into Eq. \ref{eq:rateordered}
\begin{eqnarray}
  R_\mathrm{plat} &=& (1-c_\mathrm{plat} e^{-\epsilon_w})(pc_\mathrm{plat} - e^{\epsilon_s}) \\
      &=& \frac{p e^{\epsilon_w}}{4}(1-e^{\Delta f})^2 \label{eq:plateaurate}
\end{eqnarray}
where $\Delta f= \epsilon_s-\epsilon_w-\ln p$ is the free energy difference between the red and blue states. Eq. \ref{eq:plateaurate} says that a larger free energy difference between the ordered and disordered states has a twofold effect on increasing the plateau height. First, a wider gap gives a larger linear regime for the growth rates to rise and, second, it allows for a lower level of poisoning at the plateau onset. Note that smaller values of $p$ depress the plateau height by reducing the fraction of successful additions.

{\bf Disordered particles are an increasing contribution to filament growth in the plateau.} Eq. \ref{eq:rateordered} describes an assembly rate that declines when $c$ rises above $c_\mathrm{plat}$, however, it is more common to see rates that remain constant or weakly rise. The discrepancy is that $R_\mathrm{ord}$ does not account for the incorporation of red particles in the filament. Red incorporation is accounted for by the defect capture process illustrated in Fig. \ref{FigProcesses}. The contribution of defect capture to the growth rate can be written by summing the rates of capturing defects of all sizes
\begin{eqnarray}
  R_\mathrm{def} &=& k_\mathrm{on}^\mathrm{(blue)}\sum_{n=2}^{\infty}(n+1)P_n p \\
     &=& p^2c \left(c e^{-\epsilon_w} \right)^2 \frac{3-2 c e^{-\epsilon_w}}{1-c e^{-\epsilon_w}} \label{eq:ratedefect}
\end{eqnarray}
The product $P_n p$ gives the fraction of time that disordered cap contains $n$ molecules {\em and} the terminal particle is blue. During this time a second blue molecule can attach with rate $k_\mathrm{on}^\mathrm{(blue)}=pc$, which results in a total addition of $n+1$ particles. The summation begins at $n=2$ because this is the minimum size that contains a defect and a terminal blue particle.

Two features of Eq. \ref{eq:ratedefect} are worth attention. First, $R_\mathrm{def}$ is of order $p^2$ which means that it is suppressed relative to $R_\mathrm{ord}$, which is of order $p$ (Eq. \ref{eq:rateordered}). This means that smaller values of $p$ (i.e., higher multiplicity in the disordered ensemble) will result in a lower rate of defect capture and, hence, more pronounced plateaus. Second, $R_\mathrm{def}$ has the non-physical property that it diverges as $c\rightarrow c_\mathrm{dis}$. This is because the average length of the disordered cap becomes infinite as the concentration approaches the disordered transition. This signifies a breakdown of our pseudo-equilibrium approximation because capture events occur faster than the time required for the disordered cap to reach a pseudo-equilibrium distribution of states.

The total growth rate $R_\mathrm{tot}=R_\mathrm{ord}+R_\mathrm{def}$ is obtained from Eqs. \ref{eq:rateordered} and \ref{eq:ratedefect}
\begin{equation}\label{eq:ratetotal}
 R_\mathrm{tot}= (1-c e^{-\epsilon_w})(pc - e^{\epsilon_s}) + p^2c \left(c e^{-\epsilon_w} \right)^2 \frac{3-2 c e^{-\epsilon_w}}{1-c e^{-\epsilon_w}}
\end{equation}

Fig. \ref{FigProcesses}B plots the ordered growth rate (blue) and total growth rate (black) along with the results of kinetic Monte Carlo simulations. Also shown is the fraction of particles in the filament that are red (red line). The red fraction can be computed by modifying Eq. \ref{eq:ratedefect} to obtain the growth rate due to red particles
\begin{eqnarray}
  R_\mathrm{red} &=& k_\mathrm{on}^\mathrm{(blue)}\sum_{n=2}^{\infty}(n-1)P_n p \nonumber \\
   &=& p^2c  \frac{\left(c e^{-\epsilon_w} \right)^2}{1-c e^{-\epsilon_w}} \label{eq:redrate}
\end{eqnarray}
This formula differs from  Eq. \ref{eq:ratedefect} because $R_\mathrm{def}$ describes the addition of $n+1$ particles of which the final two are blue. Therefore, Eq. \ref{eq:redrate} only accounts for the $n-1$ red particles. The red fraction, plotted in Fig. \ref{FigProcesses}, is given by
\begin{eqnarray}
  F_\mathrm{red} &=& \frac{R_\mathrm{red}}{R_\mathrm{ord}+R_\mathrm{def}} \\
   &=& \frac{p\left(c e^{-\epsilon_w} \right)^2}{\left(1-c e^{-\epsilon_w} \right)^2\left(1- \frac{ e^{\epsilon_s}}{pc} \right)+p \left(c e^{-\epsilon_w} \right)^2\left(3-2c e^{-\epsilon_w} \right)}  \label{eq:fractionred}
\end{eqnarray}
which remains low as long as $R_\mathrm{ord}>R_\mathrm{def}$, but rises rapidly as the concentration reaches $c_\mathrm{dis}$.

\begin{figure}
    \centering
    \begin{subfigure}{0.49\textwidth}
        \centering
        \includegraphics[width=\linewidth]{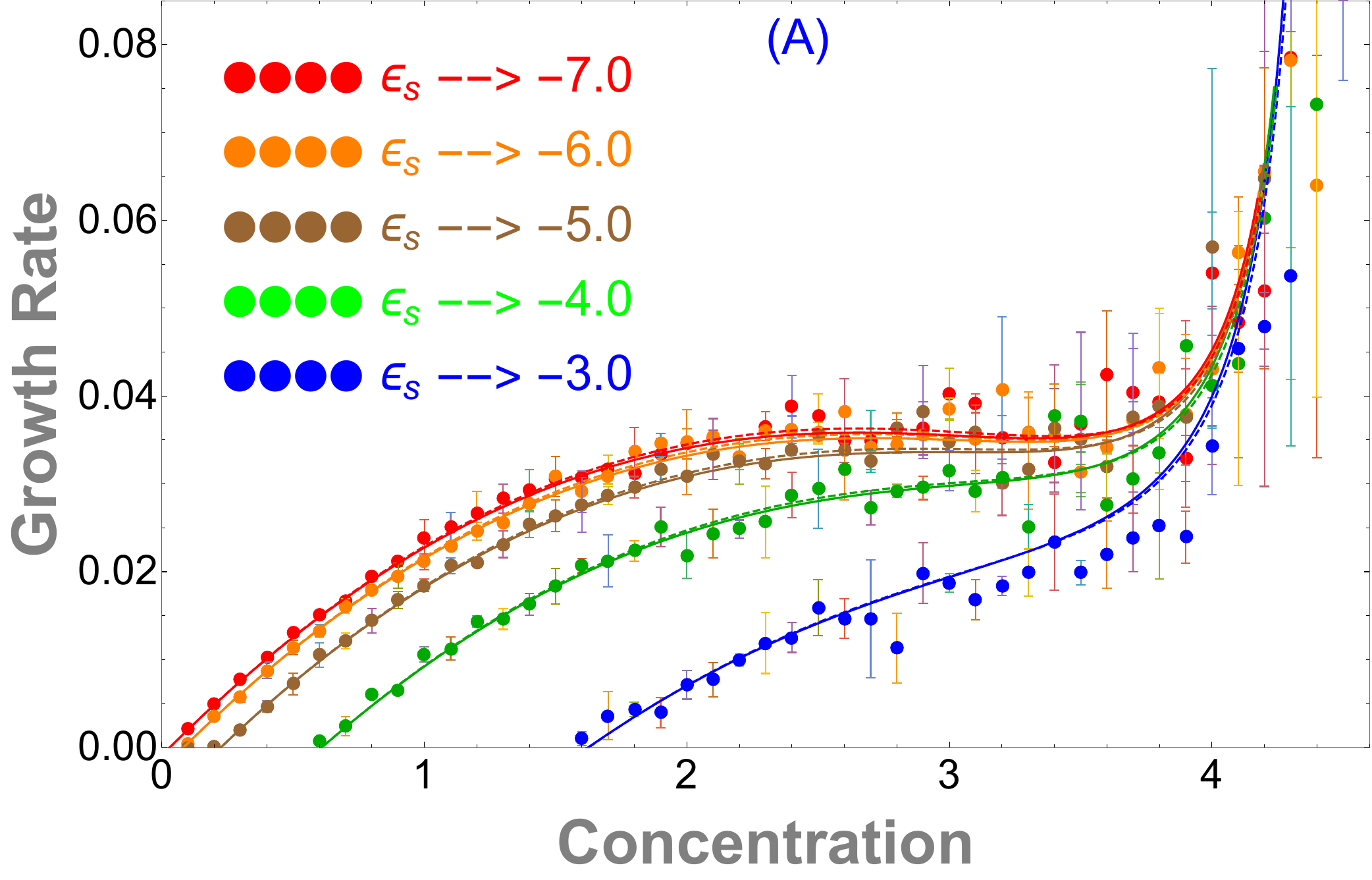}
      \label{fig:timing1}
    \end{subfigure}
    \begin{subfigure}{0.49\textwidth}
        \centering
        \includegraphics[width=\linewidth]{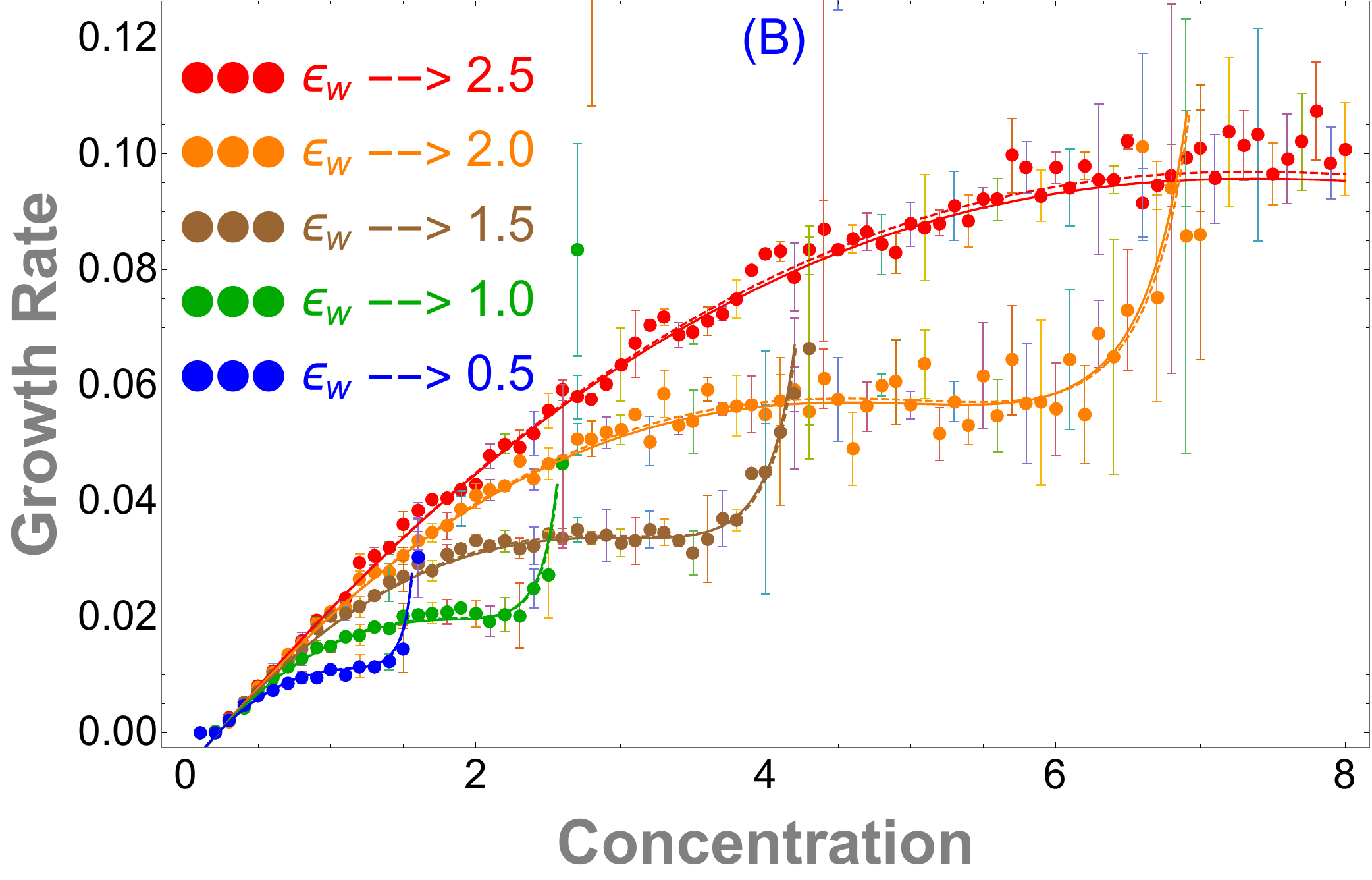}
        \label{fig:timing2}
    \end{subfigure}

    \begin{subfigure}{0.55\textwidth}
    \centering
        \includegraphics[width=\linewidth]{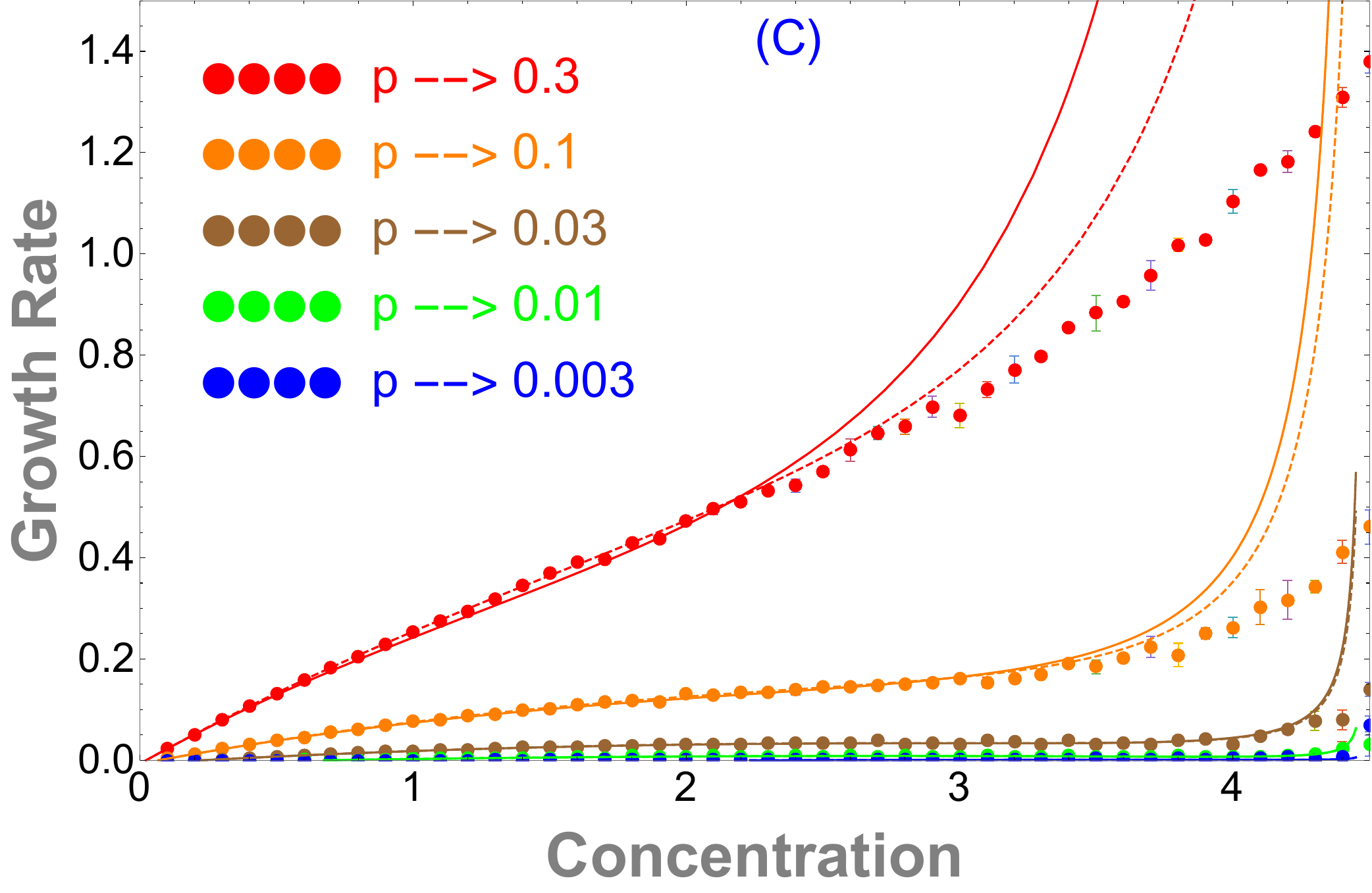}
        \label{fig:timing3}
    \end{subfigure}
\caption{{\label{fig:timing}{The shape of the growth curves is determined by the three parameters. A) More attractive values of $\epsilon_s$ shift the onset of positive growth to lower concentrations and broaden the plateau. B) More attractive values of $\epsilon_w$ shift the onset of disordered aggregation to lower concentrations and shorten the plateau. C) Smaller values of $p$ decrease the plateau height and shift the onset of ordered growth to higher concentrations. In all panels the solid lines are obtained from Eq. \ref{eq:ratetotal}, the dotted lines are the sum of Eqs. \ref{eq:rateordered2} and \ref{eq:ratedefect2}, and the dots are obtained from kinetic Monte Carlo simulations. The parameter values are $\epsilon_s = -5$, $\epsilon_w= 1.5$, $p= 0.03$ unless otherwise specified in the panel legends. }}}
\end{figure}


{\bf Pronounced plateaus are a result of high multiplicity in the disordered ensemble.} The shape of the growth curve is determined by the three model parameters: $p$, $\epsilon_s$, and $\epsilon_w$. The effects of each parameter can be seen in Fig. \ref{fig:timing}. Varying $\epsilon_s$ has the primary effect of shifting the ordered saturation concentration (Fig. \ref{fig:timing}A). Larger (less negative) values of $\epsilon_s$ shift $c_\mathrm{ord}$ to higher concentration, which causes the plateau to become narrower and less pronounced. The plateau will disappear altogether if the defect capture begins at  low enough concentration to compensate for the decline in ordered growth.

The effect of $\epsilon_w$ is seen in Fig. \ref{fig:timing}B. Making $\epsilon_w$ less attractive changes the curve in two ways. First it shifts the disorder transition, $c_\mathrm{dis}$, to higher concentration. Second, the plateau growth rate increases because there is less red binding to poison growth.

Finally, the $p$ parameter has two effects, as shown in Fig. \ref{fig:timing}C. The first is that smaller values of $p$ shift the ordered assembly solubility, $c_\mathrm{ord}$, to higher concentration. This is because greater collision rates are needed to compensate for the lower probability of ordered addition. Second, the height of the plateau is reduced because the lower probability of blue addition suppresses both $R_\mathrm{ord}$ and $R_\mathrm{def}$. Plateaus become more pronounced for small values of the ordered attachment probability and probabilities $p\lessapprox 0.1$ do not show visually distinguishable plateaus at all (see Appendix). Note that the agreement between the simulations and analytic theory degrades for larger value of $p$. This is because smaller values of $p$ allow for more particle attachment/detachment events between growth events, which makes the approximation of an equilibrated cap more accurate.

\begin{figure}
\includegraphics[width=0.95\textwidth]{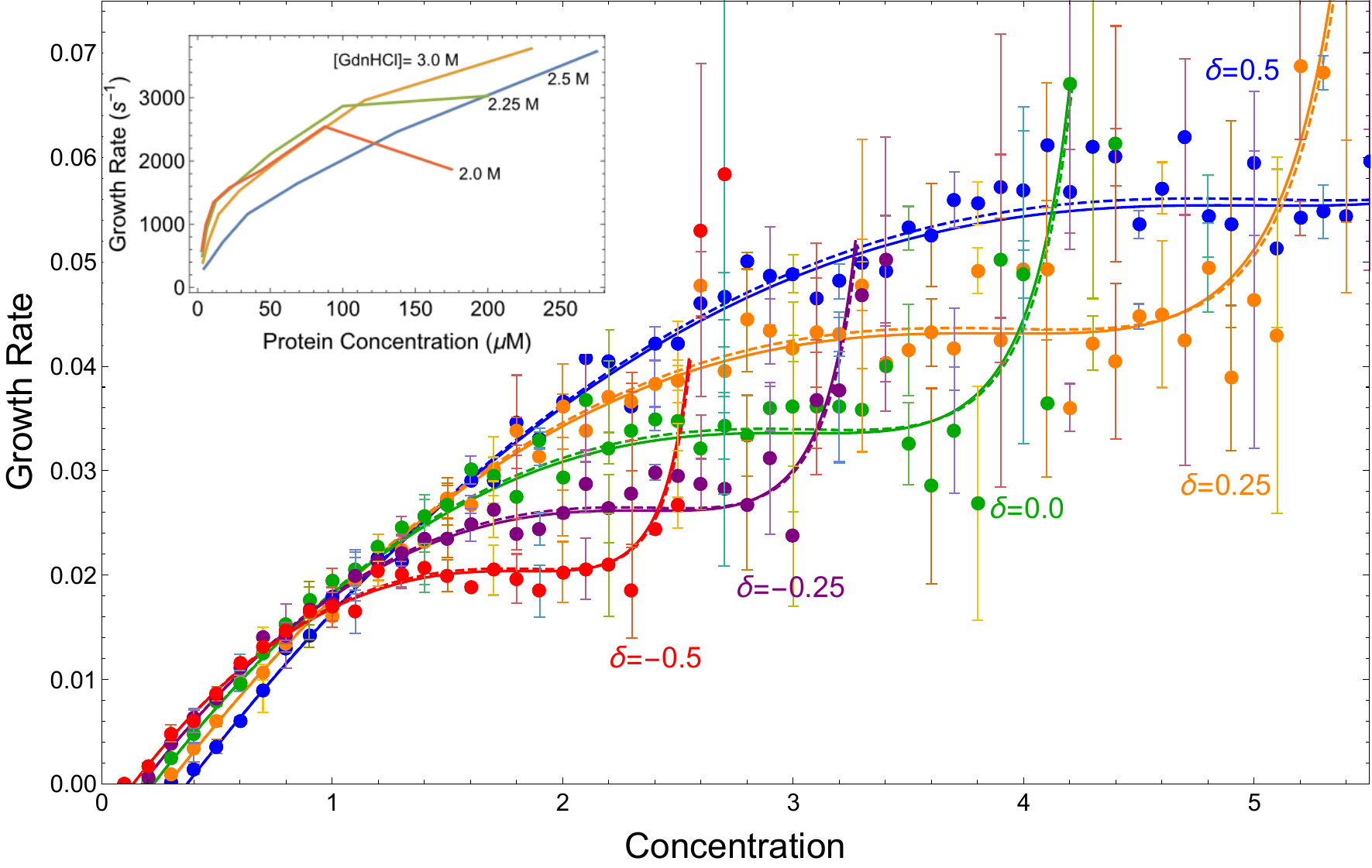}
\caption{Solution conditions that nonspecifically affect both $\epsilon_s$ and $\epsilon_w$ will have qualitatively different effects on the growth rate depending on the concentration. More attractive interactions promote growth in the linear regimes but depress growth in the plateau regime. This matches with the general trend of the experimental results of Honda et. al \cite{Honda2017} (inset). Solid lines are obtained from Eq. \ref{eq:ratetotal}, the dotted lines are the sum of Eqs. \ref{eq:rateordered2} and \ref{eq:ratedefect2}, and the dots are obtained from kinetic Monte Carlo simulations.}\label{FigNonspecific}
\end{figure}

{\bf Enhancements to binding affinity inhibit growth in the plateau regime.} One complication in using Fig. \ref{fig:timing} to interpret experimental results is that it can be difficult to independently tune the parameters. Perturbations like temperature, cosolvent additions, and mutations are expected to affect both $\epsilon_s$ and $\epsilon_w$. Fig. \ref{FigNonspecific} shows the effect on growth curves when $\epsilon_s$ and $\epsilon_w$ are simultaneously shifted in increments of 0.25 $k_BT$. Such nonspecific variation of the binding energies would be the expected result from the addition of depletants like PEG, dextran, or ficoll or denaturants like urea or guanidinium. Increasing affinity has the expected effect of shifting $c_\mathrm{ord}$ and $c_\mathrm{dis}$ to lower concentration, but the effects at intermediate concentration are more interesting. At low concentration, where disordered binding is negligible, the growth curve shifts upward due to the reduced blue detachment rate. However, the opposite effect is observed in the plateau region, where stronger binding increases growth poisoning and decreases the growth rate. Therefore, perturbations that enhance interparticle interactions can have both beneficial and inhibitory effects on assembly, depending on the concentration. This chimeric effect of nonspecific perturbations on the binding affinity has been observed in prion fibrils as the concentration of guanidinium is varied (Fig. \ref{FigNonspecific} inset). At low protein concentration low guanidinium gives the fastest growth, however, the trend reverses in the plateau regime as greater guanidinium concentrations suppress the poisoning effect of nonspecifically bound molecules.

\section{Discussion}

The difficulty in understanding self-assembly processes in biomolecular systems is that molecular anisotropy results in a large, heterogeneous ensemble of states that must be sampled. However, the size of this ensemble also means that there is necessarily a large difference between the state sampling time and the growth time \cite{Schmit2012,Schmit2013}. This separation of timescales is exploited in our model via the approximation that the non-specific ensemble reaches an equilibrium distribution between specific binding events. This approximation should be valid whenever defect incorporation events are a minority contribution to the overall growth. Within this context it makes sense that growth can be separated into the two distinct growth events, which we have called ``ordered addition'' and ``defect capture''. However, it is somewhat remarkable that the two-process approximation gives such an accurate quantitative description of the growth rate as the disorder transition is approached. This is due to the sharpness of the transition to the disordered state at $c_\mathrm{dis}$, especially for $p\ll 1$. This sharp transition is likely to be blurred in systems of real biomolecules where the non-native states will have a broad spectrum of binding affinities.

While our model lacks molecular details, it provides considerable intuition about how the binding states of a particular molecule affect the shape of the growth curve. The strongest effect comes from the characteristics of the non-native ensemble, which controls the height and width of the plateau. Furthermore, the multiplicity of the non-native ensemble even controls the slope of the linear regime where non-specific binding is negligible \cite{Larsen2023}. In contrast, the affinity of the specific state has the relatively modest effect of vertically shifting the linear, and to a lesser extent, plateau regimes.

The applicability of our model hinges on the central assumption that there is minimal conversion between non-specific and specific states while a molecule is associated with the fibril. A growing body of theoretical, experimental, and simulation work suggests this is a reasonable approximation for protein self-assembly. In particular the dominant non-specific binding states during the growth of amyloid fibrils are mis-aligned $\beta$-strands \cite{Lee2009dynamics,Schmit2013} which face a prohibitive timescale to realign while bound \cite{Roder2018} and, therefore, must unbind nearly completely to change states \cite{Jia2017,Jia2020}. Since the number of mis-aligned $\beta$-strands scales with the length of the peptide, the probability of correct binding should scale like $p\propto L^{-1}$. This relationship agrees well with the growth rate of SH3 fibrils in the linear regime \cite{Larsen2023}. One consequence of the mis-aligned states as a major component of the non-specific ensemble is that the broad contact area between the existing fibril and incoming molecule ensures that even highly localized point mutations would be expected to affect both the ordered state and the non-specific ensemble \cite{Huang2018}. For example, replacement of an amino acid with a more hydrophobic, aggregation-prone residue will enhance the binding affinity of both $\epsilon_s$ and $\epsilon_w$, leading to the chimeric enhancing/inhibitory behavior shown in Fig. \ref{FigNonspecific}. As a result, care should be taken to collect the full growth curves when assessing the effect of disease-related mutations on aggregation properties.

A central finding of this work is that the effect of a molecular or environmental perturbation depends sensitively on the concentration regime where the experiment is conducted. Perturbations that enhances intermolecular interactions increase the growth rate in the linear regime but will inhibit growth in the plateau regime by prolonging the lifetime of non-specific states. This distinction is further complicated if nucleation events are occurring because nucleation is less sensitive to non-specific binding than elongation \cite{Bunce2019,Phan2022}.

\section{Conclusion}
We have presented a model for the self-assembly of biomolecules into one-dimensional filaments. Because biomolecules have a large ensemble of non-specific states, there is necessarily a separation of timescales between state sampling and filament growth. This separation of scales allows for partitioning of growth between ordered and disordered events. Our model shows how the affinity and multiplicity of the non-specific ensemble can change the shape of the growth curve. This suggests that measurements of the growth curve can be used to obtain information about the underlying molecular mechanisms. Finally, we have shown that perturbations that affect the affinities of both the native state and the non-specific ensemble, such as depletants, osmolytes and even mutations, will have qualitatively different effects on the growth rate depending on the concentration. Therefore, it is misleading to draw conclusions about the effects of such perturbations from experiments conducted in a narrow concentration range.

{\bf Acknowledgements} This work was supported by NIH grant R01GM141235. JDS acknowledges a Beaufort Fellowship from St. John's College at the University of Cambridge.
\section{Appendix}

\textbf{Calculation of exact partition function}

The partition function for the disordered cap given by Eq. \ref{eq:capPartition} is an approximation because it does not exclude cap states that contain consecutive blue particles. These states are not allowed within the cap because a strong blue-blue bond is assumed to capture all previous particles within the stable core. The approximate Eq. \ref{eq:capPartition} works well for low values of $p$, but the exact partition function can  be calculated and yields a more accurate result. The grand partition function describes the disordered molecules bound to the last blue particle of the stable fibril core (Fig. \ref{FigProcesses}). This cap must start with a red particle because a blue particle would be considered part of the ordered core. The difficulty comes from the constraint that the cap cannot have two consecutive blue molecules at any point. As stated earlier, the probability of blue particle addition is given by $p$ and that of red particle is $q=1-p$. The grand partition function take the form
\begin{equation}\label{eq:GPFexactseries}
  Q_\mathrm{cap}^{\mathrm{exact}}   = \sum_{n=0}^{\infty} A_n {\left(c e^{-\epsilon_w} \right)}^n
\end{equation}
where $A_n$ is a combinatorial factor that accounts for the ways that blue particles can be distributed within a cap of length $n$ subject to the constraint that blue particles cannot be adjacent. This factor obeys  the recursion relation, \begin{align}
    A_n = q \: A_{n-1} + q \: p \: A_{n-2} \label{eq:recursion}
\end{align}
The first term of Eq. \ref{eq:recursion} says that cap of length $n$ that ends with a red particle can be obtained by adding a red particle to any cap of length $n-1$. The second term describes the perturbations of a disordered cap ending with a blue particle. In order to satisfy the consecutive blue constraint, it is necessary to take a cap of length $n-2$ and add a red followed by a blue, giving a  probability $qp$.

The recursion relation may be solved using the generating function, \begin{align}
    Y(x) = \sum_{n=0}^\infty x^n A_n \label{eq:generatingfunction}
\end{align}
We multiply both sides of Eq. \ref{eq:recursion} by $x^n$ and sum over all values of $n$
\begin{eqnarray}
  \sum_{n=0}^\infty x^n A_{n+2} &=& q \sum_{n=0}^\infty x^n A_{n+1} + pq \sum_{n=0}^\infty x^n A_n \\
  \frac{1}{x^2}(Y-x A_1-A_0) &=& \frac{q}{x}(Y-A_0)+pq Y
\end{eqnarray}
Applying the boundary conditions, $A_0 =1$ and  $A_1=q$ we find
 \begin{align}
    Y = \frac{1}{1-q x - p q x^2}
\end{align}
from which we can extract the combinatorial factors as follows
\begin{align}
    A_n = \frac{1}{n!} \frac{\partial^n Y}{\partial x^n}\bigg{|}_{x=0}
\end{align}
The exact grand partition function is then, \begin{align}
    Q_\mathrm{cap}^{\mathrm{exact}}   & =  \sum_{n=0}^{\infty} {\left(c e^{-\epsilon_w} \right)}^n \frac{1}{n!} \frac{\partial^n Y}{\partial x^n}\bigg{|}_{x=0}
\end{align}
Since the Maclaurin series of $f(x)$ is,\begin{align}
    f(x) = \sum_{n=0}^\infty \frac{x^n}{n!} \frac{\partial^n f}{\partial x^n} \bigg {|}_{x=0}
\end{align}
it follows that $Q_\mathrm{cap}^{\mathrm{exact}}=Y(c e^{-\epsilon_w})$ or
\begin{align}
    Q_\mathrm{cap}^{\mathrm{exact}}  & =    \frac{1}{1-(1-p) {\left(c e^{-\epsilon_w} \right)} (1+ p \: {\left(c e^{-\epsilon_w} \right)})}
\end{align}
Thus the ordered growth rate estimated earlier is more precisely given as
\begin{eqnarray}
  R_\mathrm{ord}^{\mathrm{exact}} &=& (pc - e^{\epsilon_s})/Q_{cap}^{\mathrm{exact}} \\
   &=& (pc - e^{\epsilon_s})({1-q \: c e^{-\epsilon_w} } (1+ p \: c e^{-\epsilon_w} )) \\
   &=& (pc - e^{\epsilon_s})(1-c e^{-\epsilon_w}[1-p(1-c e^{-\epsilon_w})-p^2 c e^{-\epsilon_w}]) \label{eq:rateordered2}
\end{eqnarray}
Comparison of Eqs. \ref{eq:rateordered} and \ref{eq:rateordered2} reveals that the correction (in the square brackets) has terms of order $p$ and $p^2$.

To calculate the rate of defect capture, we need to account for a cap which starts with a red molecule but ends with a blue molecule. To compute this probability we write the partition function as the sum of terms ending in blue particles and terms ending with a red particle
\begin{align}
     Q_{\mathrm {cap}}^{\mathrm {exact}} = 1 +  Q_{\mathrm {cap}}^{\mathrm {blue}} +  Q_{\mathrm {cap}}^{\mathrm {red}}
\end{align}
where the first term represents the state where no disordered particles are bound to the end of the fibril. Following Eq. \ref{eq:GPFexactseries}, we write the conditional partition functions as
\begin{eqnarray}
  Q_\mathrm{cap}^{\mathrm{blue}}  &=&  \sum_{n=0}^{\infty} A_n^\mathrm{blue} {\left(c e^{-\epsilon_w} \right)}^n \\
  Q_\mathrm{cap}^{\mathrm{red}}  &=&  \sum_{n=0}^{\infty} A_n^\mathrm{red} {\left(c e^{-\epsilon_w} \right)}^n
\end{eqnarray}
where the combinatorial factors are related by
\begin{align}
    A_{n}^{\mathrm {blue}} = A_{n-1}^{\mathrm {red}} \: p
\end{align}
Using the generating function method with the boundary conditions $A^{\mathrm {red}}_1= q$ and  $A^{\mathrm {red}}_2= q^2$, we find
\begin{equation}\label{eq:GPFexactRed}
  Q_{\mathrm {cap}}^{\mathrm {red}} = \frac{ q\: {\left(c e^{-\epsilon_w} \right)}}{1-q {\left(c e^{-\epsilon_w} \right)} (1+ p \: {\left(c e^{-\epsilon_w} \right)})}
\end{equation}

So, the grand partition function for the blue-ending cap is, \begin{align}
    Q_{\mathrm {cap}}^{\mathrm {blue}}  & = \sum_{n=2}^{\infty} A_{n}^{\mathrm {blue}} {\left(c e^{-\epsilon_w} \right)}^n
    \\ & =  p {\left(c e^{-\epsilon_w} \right)} \sum_{n=1}^{\infty} A_{n}^{\mathrm {red}} {\left(c e^{-\epsilon_w} \right)}^n
    \\ & =   \frac{p\: q\: {\left(c e^{-\epsilon_w}\right)}^2}{1-q {\left(c e^{-\epsilon_w} \right)} (1+ p \: {\left(c e^{-\epsilon_w} \right)})} \label{eq:GPFexactBlue}
\end{align}
Using Eq. \ref{eq:GPFexactBlue} and following the derivation of Eq. \ref{eq:ratedefect}
we find
\begin{eqnarray}
  R_\mathrm{def}^{\mathrm{exact}} &=& \frac{k_\mathrm{on}^\mathrm{(blue)}}{ Q_\mathrm{cap}^{\mathrm{exact}}}\sum_{n=2}^{\infty}(n+1) A_{n}^{\mathrm {blue}} {\left(c e^{-\epsilon_w} \right)}^n\\
   &=& \frac{p c} {Q_\mathrm{cap}^{\mathrm{exact}}}\frac{\partial}{\partial \left(c e^{-\epsilon_w} \right)}\sum_{n=2}^{\infty}\left(c e^{-\epsilon_w} \right)^{n+1}  A_{n}^{\mathrm {blue}} \\
   &=& \frac{p c} {Q_\mathrm{cap}^{\mathrm{exact}}}\frac{\partial}{\partial \left(c e^{-\epsilon_w} \right)} \left(c e^{-\epsilon_w} \: Q_{\mathrm {cap}}^{\mathrm {blue}} \right)
    \\
   &=& 3 p^2 q c \left(c e^{-\epsilon_w} \right)^2 + p^2 q c  \left(c e^{-\epsilon_w} \right)^3 \frac{(q + 2 p q  c e^{-\epsilon_w}  )}{{1-q c e^{-\epsilon_w} } (1+ p \: {c e^{-\epsilon_w} } )} \label{eq:ratedefect2}
\end{eqnarray}

The correction introduced by Eqs. \ref{eq:rateordered2} and \ref{eq:ratedefect2} to the approximate rates are of order $p$ and, accordingly, are only noticeable for the largest value $p=0.3$ (Fig. \ref{fig:timing}, red curve). The exact partition function predicts slightly faster growth rates at lower concentration because the approximation over-weights the capped states by including forbidden states with adjacent blue particles. Conversely, the exact partition function predicts slightly slower growth at high concentration because the approximation overestimates the probability of blue capped states. Both of these corrections are in excellent agreement with the kinetic Monte Carlo simulation, providing additional evidence for the model's assumptions that the disordered cap equilibrates on the growth timescale and that growth is the sum of ordered addition and defect capture events.

\textbf{Criteria for the appearance of a plateau}

A rough criteria for a visually distinguishable plateau is that the curvature remains negative at the plateau concentration given by Eq. \ref{eq:plateauconcentration}
\begin{equation}\label{eq:plateaucurvature}
  \left.\frac{d^2}{dc^2}(R_\mathrm{ord}+R_\mathrm{def})\right|_{c_\mathrm{plat}}=0
\end{equation}
which gives the condition
\begin{equation}
\label{eq:Pcritical}
  p = \frac{(c e^{-\epsilon_w})^3}{c e^{-\epsilon_w}(9-21c e^{-\epsilon_w}+19(c e^{-\epsilon_w})^2-6(c e^{-\epsilon_w})^3)}
\end{equation}
The range of possible values for $c_\mathrm{plat}$ is given by $c_\mathrm{dis}/2<c_\mathrm{plat}<c_\mathrm{dis}$, where the lower and upper ends of the range correspond to very strong and weak binding of the ordered state, respectively. In the strong binding limit, $c_\mathrm{plat}\simeq c_\mathrm{dis}/2$, we find that $p$ should be less than 0.1 for a plateau to appear. Eq. \ref{eq:Pcritical} is a monotonically decreasing function in the range $c_\mathrm{dis}/2<c<c_\mathrm{dis}$, so as $c_\mathrm{plat}$ shifts to larger values due to weaker $\epsilon_s$, smaller values of $p$ are required for a plateau to appear.




\end{document}